\documentclass[fleqn, 12pts]{article}
\usepackage{amsmath}
\usepackage{amssymb}
\usepackage{amsfonts}  
\usepackage{graphicx}
\usepackage{hyperref}
\usepackage{balance}
\usepackage[top=2cm, bottom=2cm, right=2.5cm, left=2.5cm]{geometry}
\usepackage[title]{appendix}
\numberwithin{equation}{section}
\usepackage{microtype}
\title{Harmonizing de Broglie-Bohm's Causal Interpretation with the Copenhagen Interpretation of Quantum Mechanics}
\author{A. N. Khondker \\
email: \href{mailto:akhondke@clarkson.edu}{akhondke@clarkson.edu} \\
Dept. of Electrical \& Computer Engineering, Clarkson University \\
Potsdam NY 13699-5720}
\date{}

\begin{document}
\maketitle
\begin{abstract}
The non-relativistic quantum theory has been interpreted causally by de Broglie, David Bohm, and others, where a quantum entity is viewed as a particle with a definite position and momentum. This interpretation opposes the Copenhagen orthodox interpretation, which expresses uncertainty through the commutator relationship between position and momentum operators. This is further exacerbated by the de Broglie-Bohm interpretation, which uses no operators corresponding to these observables. We reconcile these opposing viewpoints by introducing mathematical nonlinear operators for the momentum observable. While nonlinear non-Hermitian operators cannot be easily used as matrices in the Hilbert space, we show that they are implicitly embedded in de Broglie-Bohm's interpretation of quantum theory. 
\end{abstract}

\section{Introduction}
In quantum mechanics, physicists often avoid specific questions such as "How long does it take for an electron to tunnel through a barrier?" This question is invalid in the Copenhagen interpretation of quantum mechanics, as electrons cannot be treated as particles with a definable velocity. Various indirect approaches have been suggested to estimate tunnel time while avoiding a contradiction with Heisenberg's uncertainty principle and its orthodox interpretations. 

Numerous models are in the literature to estimate the tunnel time \cite{Buttiker_Landauer1982, Buttiker_Landauer1985, Buttiker_Landauer1986, Leavens1987, Anwar1989, Spiller1990, Hagmann1992, Barker1994, Landauer1994, Leavens1993, Xu2001}. Not surprisingly, these theories do not always agree with each other.  Some of these models \cite{Bohm1952, Hiley1988, BohmHiley1993, Steinberg1994}, are based on David Bohm's causal interpretation of the theory, which was initially rejected or ignored by many notable physicists due to its incompatibility with the quantum theory, despite electrons appearing particle-like in many experiments.  In 1927, Louis de Broglie \cite{deBroglie1960, deBroglie1964} proposed a pilot wave model, which even Max Born explored for a while. However, both abandoned their idea and adopted the approach by Bohr, Heisenberg, and others \cite{Holland1993, Durr2009, Durr2013, Barrett2019, Daumer1994}. Bohm was initially unaware of de Broglie's proposal while writing his 1952 paper. After the paper was published de Broglie \cite{deBroglie1960, deBroglie1964} joined him. J. S. Bell was influenced by Bohm's paper, which led to two seminal papers on quantum theory \cite{Bell1964, Bell1966}. Three experimentalists who verified Bell's theorem and quantum entanglement were awarded Nobel prizes in physics in 2022.

There has been considerable interest in Bohm's model in the last three decades \cite{Oriols2021}. Lopreore and Wyatt \cite{Lopreore1999}, \cite{Wyatt1999}, \cite{Wyatt_2005}  utilized the quantum trajectory method (QTM) to the dynamics of quantum wave packets, explicitly focusing on reactive scattering processes. The QTM is based on the hydrodynamic formulation of quantum mechanics, first introduced by Madelung \cite{Madelung1926} and later developed by Bohm, which treats quantum systems using trajectory-based methods. Bohmian mechanics, which offers a trajectory-based interpretation of quantum systems, has also gained popularity due to its computational and interpretational advantages. Unlike traditional quantum methods that require quantum-classical correspondence, Bohmian mechanics allows for the direct analysis of quantum systems using trajectories that evolve based on the wave function. Sanz's  \cite{Sanz2018} review paper explores the growing role of Bohmian mechanics in quantum chemistry, chemical physics, physical chemistry, molecular dynamics, and statistical mechanics. It highlights how Bohmian mechanics has been used to develop computational tools that complement traditional quantum chemistry methods. Overall, it underscores the potential of Bohmian mechanics to deepen our understanding of molecular systems and quantum dynamics, making it a valuable tool in theoretical and computational chemistry. The causal model has also gained popularity among engineers \cite{Oriols2021_2}. This paper justifies the use of de Broglie-Bohm's model by chemists, engineers, and others by providing a different approach.

\section{Overview of Bohmian Mechanics}
Louis de Broglie and Bohm treated electrons as particles with a definite position and momentum at any instant. The de Broglie-Bohm mechanics uses the Schrödinger wave equation. In Bohm's 1952 paper, the wave function is a priori expressed as:
\begin{equation}\label{eq:ref2.1}
\psi(\hat{\mathbf{r}},t) = R(\hat{\mathbf{r}},t) \exp\left(\frac{iS(\hat{\mathbf{r}},t)}{\hbar}\right)
\end{equation}
which, when substituted in the one-particle time-dependent Schrödinger's equation 
\begin{equation}\label{eq:ref2.2}
i\hbar \frac{\partial \psi(\hat{\mathbf{r}},t)}{\partial t} = -\frac{\hbar^2}{2m} \nabla^2 \psi(\hat{\mathbf{r}},t) + V(\hat{\mathbf{r}}) \psi(\hat{\mathbf{r}},t)
\end{equation}
leads to two real equations (where \( P(\hat{\mathbf{r}},t) = R^2(\hat{\mathbf{r}},t) = |\psi(\hat{\mathbf{r}},t)|^2 \)):
\begin{align}
&\frac{\partial P}{\partial t} + \nabla \cdot \left(P \frac{\nabla S}{m}\right) = 0 \label{eq:ref2.3}\\
&-\frac{\partial S}{\partial t} = \frac{(\nabla S)^2}{2m} + V(\hat{\mathbf{r}}) - \frac{\hbar^2}{2m} \frac{\nabla^2 R}{R} \label{eq:ref2.4}
\end{align}
To simplify this work, we will focus on the zero-spin, single-particle equation and omit any mention of many-particle models \cite{Bohm1952, Holland1993, Durr2013}. However, this does not detract from the physical principles at play. Equations \eqref{eq:ref2.3} and \eqref{eq:ref2.4} bear a striking resemblance to classical hydrodynamic equations \cite{Madelung1926, Hirshfelder1974}. The probability density of particles \( P(r,t) \) is conserved as expressed in Eq. \eqref{eq:ref2.3}. Bohm noted that \( S(r,t) \) in Equation (2.4) satisfies the classical Hamilton-Jacobi equation as in Eq. \eqref{eq:ref2.5} \cite{Liboff1990} if \( \hbar \rightarrow 0 \):
\begin{equation}
-\frac{\partial S}{\partial t} = H(\hat{\mathbf{r}}, \nabla S) \label{eq:ref2.5}
\end{equation}
In classical mechanics, \( S(\hat{\mathbf{r}},t) \) represents the action term and \( H(\hat{\mathbf{r}}) \) the Hamiltonian.   The action term is the time integral of the Lagrangian, which is the difference between the kinetic energy and potential \cite{Liboff1990}.  Bohm equates \( \frac{\nabla S(\hat{\mathbf{r}},t)}{m} \) as the velocity of the particle passing through the point, thus making \( \frac{(\nabla S)^2}{2m} \) the particle's kinetic energy. Furthermore, Bohm identifies the last term as a quantum mechanical potential \( V_{qu} \) when \( \hbar \neq 0 \) in addition to the 'classical' potential \( V(\hat{\mathbf{r}}) \).  The mathematical form of the Bohm model resembles a fluid-flow equation in which the flow lines cannot cross because the velocity field is a single-valued function of position \cite{Steinberg1994}. In Bohm's model, trajectories are guided by both classical and quantum forces, the latter derived from the curvature of the quantum amplitude. Interestingly, Feynman mentioned in 1948 that ``the probability amplitude has a phase proportional to the action computed classically'' for a path in space-time. He mentions that this is ``true if the action is the time integral of a quadratic function of velocity'' \cite{Feynman1948}.

In the de Broglie-Bohm interpretation, Heisenberg's commutation relation \( \left[\hat{\mathbf{r}}, \frac{\hbar \nabla}{i}\right] = i \hbar \) does not imply an inherent indeterminacy of the respective conjugate variables.  The indeterminacy is only due to the act of measurement, and the particles always have well-defined positions and momentum. To most skeptics, merely splitting up Schrödinger's equation using a wavefunction of the form given by Eq. \eqref{eq:ref2.1} to arrive at two classical mechanics-like equations and ignoring uncertainty restrictions is unacceptable. According to orthodox quantum theory, \( \frac{\nabla S(\hat{\mathbf{r}},t)}{m} \) can be considered at best some kind of “velocity” \cite{Sakurai2021}. Therefore, the term \( \frac{(\nabla S)^2}{2m} \) cannot be precisely defined as the kinetic energy of a particle.  Some have not accepted the concept of quantum potential \( V_{qu} \). Barker \cite{Barker1994} proposed a second deconstruction of Schrödinger's equation and split the quantum potential term into two parts. One part was added to the kinetic energy, and the other was kept as fluctuation energy. He also pointed out that the quantum force \( -\nabla V_{qu} \) is similar to the pressure tensor in Madelung's model \cite{Madelung1926}, which Madelung called an “internal” force of the continuum.

\section{New Momentum Operators}
Since the causal interpretation of Bohmian quantum mechanics lacks broad acceptance, it may be reasonable to explore alternative approaches to introducing objective reality. One such approach could be searching for a new operator within the de Broglie-Bohm framework, as the conventional quantum mechanical momentum operator is incompatible with the causal interpretation. However, finding a suitable “whatever” velocity operator is challenging as it must be linear to represent physical quantities. In 1945, London proposed the idea of a new velocity operator, but strictly enforced the constraint of linearity \cite{London1945}. London criticized Landau's velocity operator as merely “a formal construction devoid of physical significance” because it violated the first principles of quantum mechanics. Therefore, London cautioned against introducing arbitrary operators into the framework of quantum mechanics, as it would contradict the first principle of quantum mechanics, especially in the case of a local velocity operator at a particular point.

The difficulty arises because, in quantum mechanics, the momentum operator \( \frac{\hbar \nabla}{i} \) is a linear Hermitian operator that satisfies the fundamental postulates of the theory. Linear matrix operators are defined in a Hilbert space, while the Hermiticity requirement for quantum mechanical operators ensures that the expectation values of measurable quantities are real. Since the Hamiltonian energy operator is self-adjoint, it guarantees the conservation of probability density, leading to Eq. \eqref{eq:ref2.3}. Until the end of the 20th century, non-Hermitian operators were not widely used. However, introducing such operators with parity-time reflection (PT) symmetry with real eigenvalues has proven successful\cite{Bender1998}.

This paper explores the feasibility of using nonlinear momentum operators for the linear Schrödinger equation. Weinberg published a general framework that introduces non-linear corrections in quantum mechanics \cite{Weinberg1989}. Although nonlinear mathematical operators have been studied in various unpublished papers \cite{Rowinski2019, Jamali2022}, this study aims to demonstrate that they are potential operators that are consistent with the interpretations of quantum mechanics. Schwartz \cite{Schwartz1997} introduced and expanded upon a methodology for constructing a new mathematical framework for nonlinear operators that act on a vector space. This innovative approach eliminates the need to invoke adjoints or dual spaces. In Schwartz's words, “Whereas so much of traditional quantum theory is based upon the assumption of superposition—mandating linear operators in a vector space—it is again surprising how much can still be achieved if one abandons that habit.”

Our approach, therefore, differs from that of Madelung, de Broglie, and Bohm, who assumed the form of a wave function and used it to decompose Schrödinger's equation into two differential equations. Instead, this study proposes a new mathematical approach that can directly separate the Schrödinger Hamiltonian operator equation into two operator-like equations, resulting in a novel perspective that ultimately recovers Bohm's equations. Moreover, the study connects the origin of nonlocality in Bohm's causal model with Heisenberg's uncertainty principle of the orthodox model. We recognize that nonlinear operators are unorthodox and challenging. Nevertheless, this study illustrates how de Broglie-Bohmian mechanics implicitly adopts nonlinear mathematical functions as operators.

In this section, we will introduce a new momentum operator that satisfies the following requirements:
\begin{enumerate}
    \item \textbf{A mathematical definition:} It must have a formal definition.
    \item \textbf{Commutation Property:} It must commute with the position operator \( \mathbf{r} \) for the momentum to be truly 'local'.
    \item \textbf{The expectation value:} The expectation value of the operator must be real and provide a useful answer similar to the linear Hermitian quantum momentum operator \( \hat{\mathbf{p}}_Q = \frac{\hbar \nabla}{i} \).
\end{enumerate}

To achieve this, we start with the linear Hermitian quantum momentum operator \( \hat{\mathbf{p}}_Q = \frac{\hbar \nabla}{i} \) and rewrite its action on the \( \psi(\hat{\mathbf{r}},t) \)-function. We then propose a new momentum operator that satisfies the above requirements and briefly discuss its function and properties:
\begin{equation} \label{eq:ref3.1}
\hat{\mathbf{p}}_Q \psi(\hat{\mathbf{r}},t) = p_Q(\hat{\mathbf{r}},t) \psi(\hat{\mathbf{r}},t) = \{ \frac{\hbar}{i} \nabla \ln \psi(\hat{\mathbf{r}},t) \} \psi(\hat{\mathbf{r}},t)
\end{equation}
\textbf{Definition:} Using Eq. \eqref{eq:ref3.1}, we decompose \( \hat{\mathbf{p}}_Q \) into two mathematical operators \( \hat{\mathbf{p}}_R \) and \( i\hat{\mathbf{p}}_I \) as follows:
\begin{align}
\hat{\mathbf{p}}_R \psi(\hat{\mathbf{r}},t) & := p_R(\hat{\mathbf{r}},t) \psi(\hat{\mathbf{r}},t)   = \Re e\{\frac{\hbar}{i} \nabla \ln \psi(\hat{\mathbf{r}},t)\} \psi(\hat{\mathbf{r}},t) \label{eq:ref3.2} \\
\hat{\mathbf{p}}_I \psi(\hat{\mathbf{r}},t) & := p_I(\hat{\mathbf{r}},t) \psi(\hat{\mathbf{r}},t)  = \Im m\{\frac{\hbar}{i}  \nabla \ln \psi(\hat{\mathbf{r}},t)\} \psi(\hat{\mathbf{r}},t) \label{eq:ref3.3}
\end{align} 
Here, \( \Re e\{\} \) and \( \Im m\{\} \) are the real and imaginary parts of the functions enclosed within the curly parentheses. Thus, these mathematical operators, when operating upon any function, produce products of two real vector functions \( p_R(\hat{\mathbf{r}},t) \) and \( p_I(\hat{\mathbf{r}},t) \), respectively, with the original function. It is evident that if we add these two new mathematical operators as \( \hat{\mathbf{p}}_R + i\hat{\mathbf{p}}_I \), we recover the Hermitian momentum operator \( \hat{\mathbf{p}}_Q = \frac{\hbar \nabla}{i} \).

\textbf{Commutative properties of operators: \( \hat{\mathbf{p}}_R \) and \( \hat{\mathbf{r}} \):} To establish that the term \( \hat{\mathbf{p}}_R(\hat{\mathbf{r}},t) \) represents a definite momentum of a particle (not just 'some kind of momentum'), we need to prove the following: the local momentum and the position operators \( \hat{\mathbf{p}}_R \) and \( \hat{\mathbf{r}} \) respectively commute with each other, that is, \( [\hat{\mathbf{r}}, \hat{\mathbf{p}}_R] = 0 \).

\textbf{Proof:} Using Eq. \eqref{eq:ref3.2}, we can write
\begin{align} \label{eq:ref3.4}
(\hat{\mathbf{p}}_R \hat{\mathbf{r}}) \psi(\hat{\mathbf{r}},t)  &= \hat{\mathbf{p}}_R(\hat{\mathbf{r}} \psi(\hat{\mathbf{r}},t)) \notag \\
&= \Re e\left\{\frac{\hbar}{i}  \nabla \ln (\hat{\mathbf{r}} \psi(\hat{\mathbf{r}},t))\right\} (\hat{\mathbf{r}} \psi(\hat{\mathbf{r}},t)) \notag \\
& = \Re e\left\{\frac{\hbar}{i}  \nabla \ln \hat{\mathbf{r}} +\frac{\hbar}{i}  \nabla \ln \psi(\hat{\mathbf{r}},t)\right\}(\hat{\mathbf{r}} \psi(\hat{\mathbf{r}},t)) \notag \\
& = \hat{\mathbf{r}} \Re e\left\{\frac{\hbar}{i}  \nabla \ln \psi(\hat{\mathbf{r}},t)\right\} \psi(\hat{\mathbf{r}},t) \notag \\
& = (\hat{\mathbf{r}} \hat{\mathbf{p}}_R) \psi(\hat{\mathbf{r}},t) 
\end{align}
Here, \( \Re e\left\{\frac{\hbar}{i}  \nabla \ln \hat{\mathbf{r}} \right\} = 0 \) because the logarithm of a vector \( \hat{\mathbf{r}} \) is real. Thus, the commutator \( [\hat{\mathbf{r}}, \hat{\mathbf{p}}_R] = (\mathbf{r} \hat{\mathbf{p}}_R - \hat{\mathbf{p}}_R \hat{\mathbf{r}}) = 0 \). This means \( \Delta \hat{\mathbf{r}} \Delta \hat{\mathbf{p}}_R \geq \left|\frac{1}{2} [\hat{\mathbf{r}}, \hat{\mathbf{p}}_R]\right| = 0 \).
We emphasize that the above proof does not imply the ability to measure a particle's position and momentum without any unpredictable uncertainty. However, it strongly suggests that it is possible to assign both momentum and position to a point particle simultaneously without violating any principles if one recognizes the mathematical rule \( \hat{\mathbf{p}}_R \) as the valid momentum operator in the Bohm formalism and \( p_R(\hat{\mathbf{r}},t) \) is the local momentum of the particle.

\textbf{The expectation value of \( \hat{\mathbf{p}}_R \):} Using the fact that the expectation value of the Hermitian momentum operator \( \hat{\mathbf{p}}_Q \) is always real, one obtains the following:
\begin{align} \label{eq:ref3.5}
\langle \hat{\mathbf{p}}_Q \rangle  &= \langle \psi(\hat{\mathbf{r}},t) | \hat{\mathbf{p}}_Q | \psi(\hat{\mathbf{r}},t) \rangle \notag \\
&= \langle \psi(\hat{\mathbf{r}},t) | \hat{\mathbf{p}}_R + i\hat{\mathbf{p}}_I | \psi(\hat{\mathbf{r}},t) \rangle \notag \\
&= \langle \psi(\hat{\mathbf{r}},t) | \hat{\mathbf{p}}_R | \psi(\hat{\mathbf{r}},t) \rangle + i \int d\hat{\mathbf{r}} \ \psi^*(\hat{\mathbf{r}},t) \hat{\mathbf{p}}_I \psi(\hat{\mathbf{r}},t) \notag \\
&= \int d\hat{\mathbf{r}} \ \psi^*(\hat{\mathbf{r}},t) p_R \psi(\hat{\mathbf{r}},t) + i \int d\hat{\mathbf{r}}  \, \psi^*(r,t) p_I \psi(\hat{\mathbf{r}},t) 
\end{align}
Here, \( d\hat{\mathbf{r}} \) represents a volume integration. Since functions \( p_R(\hat{\mathbf{r}},t) \) and \( p_I(\hat{\mathbf{r}},t) \) are real, the second imaginary term must be dropped for physical quantities, e.g., momentum. We conclude
\begin{align} \label{eq:ref3.6}
\langle \hat{\mathbf{p}}_Q \rangle  &= \int d\hat{\mathbf{r}} \, |\psi(\hat{\mathbf{r}},t)|^2 p_R(\hat{\mathbf{r}},t) \notag \\ 
&= \langle \psi(\hat{\mathbf{r}},t) | \hat{\mathbf{p}}_R | \psi(\hat{\mathbf{r}},t) \rangle   \notag \\
&= \langle \hat{\mathbf{p}}_R \rangle
\end{align}

In other words, the new mathematical momentum operator \( \hat{\mathbf{p}}_R \) must produce the same expectation value as the linear Hermitian momentum operator \( \hat{\mathbf{p}}_Q \). However, Eq. \eqref{eq:ref3.6} has a new meaning. The average momentum is a statistical ensemble of possible momenta \( p_R(\hat{\mathbf{r}},t) \)of a particle with a probability density \( |\psi(\hat{\mathbf{r}},t)|^2 \) within a volume of interest. The last statement can also be understood in light of the “problem of stationary states” \cite{Bohm1952} and the quantum equilibrium hypothesis \cite{Bohm1953}, \cite{Holland1993}, \cite{Goldstein2007}. We mention that while in the orthodox interpretation, \( d\hat{\mathbf{r}} \, |\psi(\hat{\mathbf{r}},t)|^2 \) represents the probability of finding a particle within a volume element \( d\hat{\mathbf{r}} \) when a measurement is made. In the causal interpretation, it denotes the probability of a particle being at position \(\hat{\mathbf{r}}\) at time \( t \), even if no measurement is made.

\textbf{The role of \( \hat{\mathbf{p}}_I \) in quantum mechanics:} The function of \( \hat{\mathbf{p}}_I \) in Bohmian mechanics is as follows: As \( \hat{\mathbf{p}}_Q = \hat{\mathbf{p}}_R + i\hat{\mathbf{p}}_I \), we can use the Heisenberg commutator relationship \( [\hat{\mathbf{r}}, \hat{\mathbf{p}}_Q] = i \hbar \) along with Eq. \eqref{eq:ref3.4} to show that the commutator relationship \( [\hat{\mathbf{r}}, i\hat{\mathbf{p}}_I] = i \hbar \). Additionally, since \( \langle \hat{\mathbf{p}}_Q \rangle = \langle \hat{\mathbf{p}}_R \rangle \), the expectation value \( \langle i\hat{\mathbf{p}}_I  \rangle = 0 \). The most important observation of this is that the operator \( i\hat{\mathbf{p}}_I \) does not contribute to the local momentum of the particle. Yet it is solely responsible for Heisenberg's non-commuting relationship, which creates a peculiar fuzziness with the particle's location and momentum in quantum theory. In section 5, we will elaborate on how \( i\hat{\mathbf{p}}_I\) plays a crucial role in both interpretations of quantum mechanics.

To complete the de Broglie-Bohm picture, we use the results of Eq. \eqref{eq:ref3.2} to derive its guiding equation by defining the local velocity of a particle as
\begin{align} \label{eq:ref3.7}
\hat{\mathbf{v}}_r(\hat{\mathbf{r}},t) & = \frac{\partial \hat{\mathbf{r}}(t)}{\partial t} \notag \\
& = \frac{p_R(\hat{\mathbf{r}},t)}{m} \notag \\
& = \Re e\left(\frac{\hbar}{im} \nabla \ln \psi(\hat{\mathbf{r}},t)\right)
\end{align}
Equation \eqref{eq:ref3.7} can also be rewritten as       
\begin{align} \label{eq:ref3.8}
\hat{\mathbf{v}}_r(\hat{\mathbf{r}},t) & = \frac{\hbar}{m} \Im m\left(\frac{\psi^*(\hat{\mathbf{r}},t) \nabla \psi(\hat{\mathbf{r}},t)}{|\psi(\hat{\mathbf{r}},t)|^2} \right) 
= \frac{\hat{\mathbf{J}}(\hat{\mathbf{r}},t)}{|\psi(\hat{\mathbf{r}},t)|^2}
\end{align}
where \( \hat{\mathbf{J}}(\hat{\mathbf{r}},t) \) is the probability current density of the wave function.

\section{The Partial Time Derivative Term in Schrödinger's Equation}
Unlike a particle's momentum, position, and energy, time \( t \) is not an observable quantity in non-relativistic closed quantum mechanics. The partial time derivative in Schrödinger's equation is not a Hermitian (self-adjoint) operator. Being mindful of this, we nevertheless introduce two new mathematical rules \( e_R(\hat{\mathbf{r}},t) \) and \( e_I(\hat{\mathbf{r}},t) \) respectively, as follows:
\begin{align}
e_R(\hat{\mathbf{r}},t) \psi(\hat{\mathbf{r}},t) & \equiv \Re e\left\{ i \hbar \frac{\partial}{\partial t} \ln \psi(\hat{\mathbf{r}},t) \right\} \psi(\hat{\mathbf{r}},t) \label{eq:ref4.1}\\
e_I(\hat{\mathbf{r}},t) \psi(\hat{\mathbf{r}},t) & \equiv \Im m\left\{ i \hbar \frac{\partial}{\partial t} \ln \psi(\hat{\mathbf{r}},t) \right\} \psi(\hat{\mathbf{r}},t) \label{eq:ref4.2}
\end{align}
It is easy to verify that
\begin{equation}
e_R(\hat{\mathbf{r}},t) \psi(\hat{\mathbf{r}},t) + i e_I(\hat{\mathbf{r}},t) \psi(\hat{\mathbf{r}},t) = i \hbar \frac{\partial \psi(\hat{\mathbf{r}},t)}{\partial t} \label{eq:ref4.3}
\end{equation}
We will use the two mathematical rules to derive two coupled operator equations for the Bohmian mechanics. Appendix A shows a few properties of these two mathematical rules, which will be used in the next section.

\section{Bohm-de Broglie Equations Revisited}
It is known that a rigorous “derivation” of Schrödinger's equation does not exist; instead, it is postulated. One may begin with the classical energy equation, i.e., the total energy is the sum of kinetic and potential energy:
\begin{equation} \label{eq:ref5.1}
E = \frac{p_{\text{classical}}^2}{2m} + V(\hat{\mathbf{r}}) 
\end{equation}
and replace each classical term for momentum, position, and energy (\( p_{\text{classical}} \), \( \hat{\mathbf{r}} \), and \( E \), respectively) with its corresponding quantum mechanical operator. In other words, \( p_{\text{classical}} \rightarrow \frac{\hbar \nabla}{i} = \hat{\mathbf{p}}_Q \), \( \hat{\mathbf{r}} \rightarrow \hat{\mathbf{r}} \), and \( E \rightarrow i \hbar \frac{\partial}{\partial t} \), which results in an equation involving only operators and a time derivative:
\begin{equation} \label{eq:ref5.2}
i \hbar \frac{\partial}{\partial t} = \frac{\hat{\mathbf{p}}_Q \cdot \hat{\mathbf{p}}_Q}{2m} + V(\hat{\mathbf{r}}) 
\end{equation}
When Eq. \eqref{eq:ref5.2} acts on the \( \psi(\hat{\mathbf{r}},t) \)-function, it results in the well-known Schrödinger's wave equation, Eq \eqref{eq:ref2.2}. The right-hand side of Eq. (\eqref{eq:ref5.2} represents the Hermitian Hamiltonian Energy operator. This operator has the property of preserving the continuity of the probability current density. Thus, it is not surprising that Schrödinger's equation embodies both the principles of particle number conservation and energy conservation.

When we substitute \( \hat{\mathbf{p}}_Q \rightarrow \hat{\mathbf{p}}_R + i\hat{\mathbf{p}}_I \) and \( i \hbar \frac{\partial}{\partial t} \rightarrow e_R(t) + i e_I(t) \) in the above, we obtain
\begin{align}
&e_R(t) + i e_I(t) = \frac{\hat{\mathbf{p}}_R \cdot \hat{\mathbf{p}}_R}{2m} + V(\hat{\mathbf{r}}) - \frac{\hat{\mathbf{p}}_I \cdot \hat{\mathbf{p}}_I}{2m}  + \frac{i}{2m}(\hat{\mathbf{p}}_R \cdot \hat{\mathbf{p}}_I + \hat{\mathbf{p}}_I \cdot \hat{\mathbf{p}}_R)   \label{eq:ref5.2a} 
\end{align}
We separate the real and the imaginary parts Eq. \eqref{eq:ref5.2a} into two equations, i.e.,
\begin{align}
&e_I(t) = \frac{1}{2m}(\hat{\mathbf{p}}_R \cdot \hat{\mathbf{p}}_I + \hat{\mathbf{p}}_I \cdot \hat{\mathbf{p}}_R) \label{eq:ref5.3}\\
&e_R(t) = \frac{\hat{\mathbf{p}}_R \cdot \hat{\mathbf{p}}_R}{2m} + V(\hat{\mathbf{r}}) + \frac{i\hat{\mathbf{p}}_I \cdot i\hat{\mathbf{p}}_I}{2m} \label{eq:ref5.4}
\end{align}

Equations \eqref{eq:ref5.3} and \eqref{eq:ref5.4} are two coupled equations that replace the single operator equation \eqref{eq:ref5.2}. It is worth noting that these two equations are independent of any specific form of the \( \psi(\hat{\mathbf{r}},t) \)-function. Let's focus on Eq. \eqref{eq:ref5.4}, where \( \hat{\mathbf{p}}_R \) represents the local momentum operator of the particle. Using the conclusion of Eq. (A.3), one can postulate that the local kinetic energy operator of the particle is given by \( \frac{\hat{\mathbf{p}}_R \cdot \hat{\mathbf{p}}_R}{2m} \). The term \( V(\hat{\mathbf{r}}) \) is the mathematical operator for the classical potential energy of the particle. Therefore, it is reasonable to assume that the last term \( \frac{i\hat{\mathbf{p}}_I \cdot i\hat{\mathbf{p}}_I}{2m} \) represents some form of a potential energy operator.  

Now we provide a connection to the de Broglie-Bohm model if we use \( \psi(\hat{\mathbf{r}},t) \)-function represented by equation (\eqref{eq:ref2.1}, to derive the following two equations. 
\begin{align}
&\int_{-\infty}^{\infty} d\hat{\mathbf{r}} \, |\psi(\hat{\mathbf{r}},t)|^2 \left[ \frac{\partial P}{\partial t} + \nabla \cdot \left(P \frac{\nabla S}{m}\right)\right] =  0 \label{eq:ref5.5}\\
&\int_{-\infty}^{\infty} d\hat{\mathbf{r}} \, |\psi(\hat{\mathbf{r}},t)|^2  \left[-\frac{\partial S}{\partial t} \right] =  \int d\hat{\mathbf{r}} \, |\psi(\hat{\mathbf{r}},t)|^2 \left[\frac{(\nabla S)^2}{2m} + V(\hat{\mathbf{r}}) - \frac{\hbar^2}{2m} \frac{\nabla^2 R}{R}\right] \label{eq:ref5.6}
\end{align}
 Equations \eqref{eq:ref5.5} and \eqref{eq:ref5.6}  are reformulations of Eqns. \eqref{eq:ref2.3} and \eqref{eq:ref2.4}. Details of the derivations are shown in Appendices A and B. We have proved that Eqs. \eqref{eq:ref5.3} and \eqref{eq:ref5.4}  are the operator versions of the continuity equations and the Hamilton-Jacobian equation in Bohmian mechanics.   

By comparing equations \eqref{eq:ref2.4}, \eqref{eq:ref5.4} and \eqref{eq:ref5.6}, we can observe that the Bohmian quantum potential \( V_{qu} \) originates from the operator \( i\hat{\mathbf{p}}_I \). Our findings demonstrate that the Bohm model does not neglect the imaginary part of the neoclassical nonlinear momentum operator, as mentioned by Jamali \cite{Jamali2022}. While Jamali identified the importance of the imaginary part of the momentum operator, we have provided a clear explanation of its role within the framework of de Broglie-Bohm mechanics and the orthodox interpretation. Specifically, the operator \( \frac {i \hat{\mathbf{p}}_I \cdot i \hat{\mathbf{p}}_I}{2m} \) extracts the quantum potential \( V_{qu} \) from the \( \psi(\hat{\mathbf{r}},t) \)-function. The quantum potential is responsible for the nonlocality observed in Bohmian mechanics. This can be seen in the many-body version of \( V_{qu} \), which considers the forces between particles in different positions. This is illustrated below:
\begin{equation} \label{eq:ref5.7}
V_{qu}(\hat{\mathbf{r}}_1, \ldots, \hat{\mathbf{r}}_N, t) = -\frac{\hbar^2}{2m} \sum_{k=1}^{N}\frac{\nabla_k^2 R(\hat{\mathbf{r}}_1, \ldots, \hat{\mathbf{r}}_N, t)}{R(\hat{\mathbf{r}}_1, \ldots, \hat{\mathbf{r}}_N, t)} 
\end{equation}
The quantum potential in Bohmian mechanics demonstrates nonlocality because it depends on the entire wave function, which spans across space. This means that the motion of one particle at a single location will be instantaneously influenced by others at distant locations, thereby defying the speed limitation imposed by the special theory of relativity.

\section{Conclusions}

This paper demonstrated that nonlinear operators are essential to the Bohmian interpretation of non-relativistic quantum mechanics. We have shown that two hidden non-linear momentum operators, represented by the nonlinear momentum operators \( \hat{\mathbf{p}}_R \) and \( i\hat{\mathbf{p}}_I \), respectively, operate covertly in the Hermitian quantum mechanical operator \( \hat{\mathbf{p}}_Q \). The real momentum operator commutes with the position, justifying the consideration of an electron as a “point” particle with a local momentum.   On the other hand, Heisenberg's uncertainty relationship is demonstrated to be caused by the imaginary momentum operator \( i\hat{\mathbf{p}}_I \). Unexpectedly, this operator also appears in the quantum potential, \( -\frac{\hat{\mathbf{p}}_I \cdot \hat{\mathbf{p}}_I}{2m} \), in Bohmian mechanics. Therefore, it is unsurprising that the Bohmian approach replicates every outcome of conventional quantum physics, making them both sides of the same coin.

The wave functions in Bohmian mechanics provide the system with rules governing the motion of a point particle only guided by the probability current density, \( \hat{\mathbf{J}}(\hat{\mathbf{r}},t) = |\psi(\hat{\mathbf{r}},t)|^2 \hat{\mathbf{v}}_r(\hat{\mathbf{r}},t)\) à la Eq. \eqref{eq:ref3.8}. An electron in a closed system follows the system's rules, governed by a quantum Hamilton-Jacobi equation and continuity equations. Thus, a handy formula that encapsulates these equations is the Schrödinger equation. The linear Hermitian operators of orthodox quantum theory provide a convenient method for calculating the ensemble average of various experimental results using elegant Hilbert space representations, eigenvalues, and eigenfunctions. However, the theory is not equipped to answer questions like “Through which slit did the electron pass?” after it appears as a point dot on a photographic plate. If we consider the electron as a particle, the only way to answer such questions is through the causal interpretation of David Bohm without any hesitation.

In the last few decades, researchers have proposed several extensions of de Broglie-Bohm mechanics for closed quantum systems to incorporate spins \cite{Norsen2014}, periodic crystalline materials \cite{Deng2018}, and quantum field theory \cite{Durr2009QFT, Durr2013}. We suggest that the nonlinear momentum operator and de Broglie-Bohm approach can also be applied to open quantum systems. One such example is the nonequilibrium retarded Green's function methods of quantum kinetic (Keldysh) theory \cite{McLennan1991, Khondker1992, Alam1992, Haque1995}. In quantum kinetic theory, retarded Green's functions are employed for the Schrödinger equation:
\begin{align} \label{eq:ref6.1}
G_R(\hat{\mathbf{r}},\hat{\mathbf{r}}';E) & = |G_R(\hat{\mathbf{r}},\hat{\mathbf{r}}';E)| \exp\left(i\theta(\hat{\mathbf{r}},\hat{\mathbf{r}}';E)\right)
\end{align}

This paper does not focus on providing details of the quantum kinetic theory. However, we can briefly summarize its physical interpretation within the context of quantum transport, which involves the creation and annihilation of particles within the de Broglie-Bohm framework. Previous work proposed a semiclassical interpretation of Equation \eqref{eq:ref6.1} (which we do not reproduce here) without invoking the causal interpretation. In the De Broglie-Bohm picture, if electrons with energy \( E \) have been injected from all energy states at locations \( \hat{\mathbf{r}}' \) and a fraction of them propagate coherently, without getting scattered out of the energy state, to location \( \hat{\mathbf{r}} \), the probability of finding them at \( \hat{\mathbf{r}} \) with energy \( E \) is proportional to \( \frac{|G_R(\hat{\mathbf{r}},\hat{\mathbf{r}}';E)|^2}{2\pi} \). The total electron current density \( J_e(\hat{\mathbf{r}};E) \) per unit energy contributed by all electrons at \( \hat{\mathbf{r}} \) can be obtained by summing over all possible \( \hat{\mathbf{r}}' \) as described in previous works \cite{McLennan1991, Alam1992}
\begin{align} \label{eq:ref6.2}
J_e(\hat{\mathbf{r}};E) & = \frac{e}{2\pi} \int \frac { d\hat{\mathbf{r}}'}{\tau_p(\hat{\mathbf{r}}';E)} \, \frac{\hbar}{ m} \nabla \theta(\hat{\mathbf{r}},\hat{\mathbf{r}}';E)|G_R(\hat{\mathbf{r}},\hat{\mathbf{r}}';E)|^2
\end{align}
where \( \frac{1}{\tau_p(\hat{\mathbf{r}}'; E)} \) is the electron in-scattering rate (or hole out-scattering rate) to an energy state with energy \( E \) at \( \hat{\mathbf{r}}' \). The velocity of electrons is \( \frac{\hbar \nabla \theta(\hat{\mathbf{r}},\hat{\mathbf{r}}';E)}{m} \), which is similar to \( \frac{\nabla S(\hat{\mathbf{r}},t)}{m} \) in Bohm's approach. We believe the de Broglie-Bohm approach can naturally provide a meaningful interpretation of the quantum kinetic theory if electrons are treated as particles.

\begin{appendices}
\counterwithin*{equation}{section}
\renewcommand\theequation{\thesection.\arabic{equation}}

\section{}

\setcounter{equation}{0}
Note, in a closed conservative system, the mean particle energy is the expectation value of the Hamiltonian expressed as \cite{Bohm1952}:
\begin{align} \label{eq:refA.1}
& \langle E \rangle_{\text{ensemble average}}=\langle H \rangle   
= \int_{-\infty}^{\infty} d\hat{\mathbf{r}} \, \psi^*(\hat{\mathbf{r}},t) 
\left(-\frac{\hbar^2}{2m} \nabla^2 + V(\hat{\mathbf{r}})\right) \psi(\hat{\mathbf{r}},t)
\end{align}

Using Eq. \eqref{eq:ref2.2}, we can evaluate the total energy of the particle using the partial time-derivative of the wavefunction \( \psi(r,t) \) as follows:
\begin{align} \label{eq:refA.2}
\langle E \rangle_{\text{ensemble average}} & = \int_{-\infty}^{\infty} d\hat{\mathbf{r}} \, \psi^*(\hat{\mathbf{r}},t)  \left(i\hbar \frac{\partial \psi(\hat{\mathbf{r}},t)}{\partial t}\right)
\end{align}

Since \( \Re e\left\{i \hbar \frac{\partial}{\partial t} \ln \psi(\hat{\mathbf{r}},t)\right\} \) and \( \Im m\left\{i \hbar \frac{\partial}{\partial t} \ln \psi(\hat{\mathbf{r}},t)\right\} \) are real functions and the ensemble energy \( E \) is real, we use Eqns. \eqref{eq:ref3.1}-\eqref{eq:ref3.3} to arrive at:
\begin{align} \label{eq:refA.3}
\langle E \rangle_{\text{ensemble average}} & = \int_{-\infty}^{\infty} d\hat{\mathbf{r}} \, \psi^*(\hat{\mathbf{r}},t)  e_R(\hat{\mathbf{r}},t) \psi(r,t) = \langle e_R \rangle
\end{align}

In other words, if the wave function is known, the mathematical operator \( e_R(t) \) can be used to estimate the total ensemble average energy of electrons.

\section{}

\setcounter{equation}{0}
In Appendix B, we derive the relations shown in Eqns. \eqref{eq:ref5.5} and \eqref{eq:ref5.6}. The complex \( \psi(\hat{\mathbf{r}},t) \)-function is given as \( \psi(\hat{\mathbf{r}},t) = R(\hat{\mathbf{r}},t) \exp\left(\frac{i S(\hat{\mathbf{r}},t)}{\hbar}\right) \) as shown in \eqref{eq:ref2.1} . We begin with the linear Hermitian momentum operator, which gives 
\begin{align}
\hat{\mathbf{p}}_Q \psi(\hat{\mathbf{r}},t) & = -i \hbar \nabla \psi(\hat{\mathbf{r}},t) \notag \\
& = \left(R \nabla S - i \hbar \nabla R\right) \exp\left(\frac{i S(\hat{\mathbf{r}},t)}{\hbar}\right) \label{eq:refB.1}
\end{align}
For convenience, we define two vector functions
\begin{align}
R_1 & = R \nabla S \label{eq:refB.2}\\
R_2 & = -i \hbar \nabla R \label{eq:refB.3}
\end{align}

Since \( \hat{\mathbf{p}}_Q = \hat{\mathbf{p}}_R + i\hat{\mathbf{p}}_I \), we separate Eq. \eqref{eq:refB.1} into two equations, as follows:
\begin{align}
\hat{\mathbf{p}}_R \psi(\hat{\mathbf{r}},t) &= p_R(\hat{\mathbf{r}},t)\psi(\hat{\mathbf{r}},t)  = \nabla S \psi(\hat{\mathbf{r}},t) = R_1 \exp\left(\frac{i S(\hat{\mathbf{r}},t)}{\hbar}\right) \label{eq:refB.4}\\
i\hat{\mathbf{p}}_I \psi(\hat{\mathbf{r}},t) &= -i \hbar \nabla R \exp\left(\frac{i S(\hat{\mathbf{r}},t)}{\hbar}\right)  = R_2 \exp\left(\frac{i S(\hat{\mathbf{r}},t)}{\hbar}\right) \label{eq:refB.5}
\end{align}

Eq. \eqref{eq:refB.2} shows the local momentum \( p_R(\hat{\mathbf{r}},t) = \nabla S \). Next, we operate \( \hat{\mathbf{p}}_Q \) on Eq. \eqref{eq:refB.4}, where we use the vector dot product and use \eqref{eq:refB.1} and \eqref{eq:refB.2} :
\begin{align}
\hat{\mathbf{p}}_Q \cdot \hat{\mathbf{p}}_R \psi(\hat{\mathbf{r}},t) 
&= \left(R_1 \cdot \nabla S - i \hbar \nabla \cdot R_1 \right) \exp\left(\frac{i S(\hat{\mathbf{r}},t)}{\hbar}\right) \notag \\
& = \left[ R \nabla S \cdot \nabla S  - i \hbar \left(\nabla R \cdot \nabla S + R \nabla^2 S \right) \right] \exp\left(\frac{i S(\hat{\mathbf{r}},t)}{\hbar}\right) \notag \\
& = \left[ \left(\nabla S \cdot \nabla S \right) - i \hbar \left(\frac{\nabla R \cdot \nabla S}{R} + \nabla^2 S \right)\right] R \exp\left(\frac{i S(\hat{\mathbf{r}},t)}{\hbar}\right) \label{eq:refB.6}
\end{align}

Since \( \hat{\mathbf{p}}_Q \cdot \hat{\mathbf{p}}_R = (\hat{\mathbf{p}}_R + i\hat{\mathbf{p}}_I) \cdot \hat{\mathbf{p}}_R = \left(\hat{\mathbf{p}}_R \cdot \hat{\mathbf{p}}_R + i\hat{\mathbf{p}}_I \cdot \hat{\mathbf{p}}_R\right) \), we arrive at:
\begin{align} \label{eq:refB.7}
(\hat{\mathbf{p}}_R \cdot \hat{\mathbf{p}}_R + i\hat{\mathbf{p}}_I \cdot \hat{\mathbf{p}}_R )\psi(\hat{\mathbf{r}},t) & = \left[ \left(\nabla S \right)^2 
- i \hbar \left(\frac{\nabla R \cdot \nabla S}{R} + \nabla^2 S \right)  \right] \psi(\hat{\mathbf{r}},t)
\end{align}

Similarly, we operate \( \hat{\mathbf{p}}_Q \) on Eq. \eqref{eq:refB.5} where we use vectors dot operation and use \eqref{eq:refB.1} and \eqref{eq:refB.3}:
\begin{align}
 \hat{\mathbf{p}}_Q \cdot (i\hat{\mathbf{p}}_I )\psi(r,t) \notag 
 &=\left(R_2 \cdot \nabla S - i \hbar \nabla \cdot R_2 \right) \exp\left(\frac{i S(\hat{\mathbf{r}},t)}{\hbar}\right) \notag \\
 &= \left[ (- i \hbar \nabla R) \cdot \nabla S -i \hbar \nabla \cdot (- i \hbar \nabla R)) \right]  \exp\left(\frac{i S(\hat{\mathbf{r}},t)}{\hbar}\right) \notag \\
 & = (-\hbar^2 \frac{\nabla^2 R} {R}  
- i \hbar \frac{\nabla R \cdot \nabla S}{R} ) R \exp\left(\frac{i S(\hat{\mathbf{r}},t)}{\hbar}\right) \label{eq:refB.8}
\end{align}

Since \( \hat{\mathbf{p}}_Q \cdot i\hat{\mathbf{p}}_I = (\hat{\mathbf{p}}_R + i\hat{\mathbf{p}}_I) \cdot i\hat{\mathbf{p}}_I = \left(-\hat{\mathbf{p}}_I \cdot \hat{\mathbf{p}}_I + i\hat{\mathbf{p}}_R \cdot \hat{\mathbf{p}}_I \right) \), we arrive at:
\begin{align} \label{eq:refB.9}
\left( i\hat{\mathbf{p}}_I \cdot i\hat{\mathbf{p}}_I + i\hat{\mathbf{p}}_R \cdot \hat{\mathbf{p}}_I \right) \psi(\hat{\mathbf{r}},t)  
= \left(  -\hbar^2 \frac{\nabla^2 R }{R} - i \hbar \frac{ \nabla R \cdot \nabla S}{R} \right) \psi(\hat{\mathbf{r}},t)
\end{align}

Using Eqns. \eqref{eq:ref4.1}-\eqref{eq:ref4.3} and \eqref{eq:ref2.1}, the time-derivative term on the left side of the Schrödinger equation, as shown in Eq. \eqref{eq:ref2.2}, can be written as:
\begin{align}
i \hbar \frac{\partial \psi(\hat{\mathbf{r}},t)}{\partial t} & = i \hbar \frac{\partial}{\partial t}\left[R(\hat{\mathbf{r}},t) + i \frac{S(\hat{\mathbf{r}},t)}{\hbar}\right] \psi(\hat{\mathbf{r}},t) \notag
\end{align}
Thus: \label{eq:refB.10}
\begin{align}
[e_R(t) + i e_I(t)] \psi(\hat{\mathbf{r}},t) & = \left[-\frac{\partial S}{\partial t} + i \hbar \frac{\partial R / \partial t}{R} \right]  \psi(\hat{\mathbf{r}},t)
\end{align}
We now use the terms in equations \eqref{eq:ref5.3} and \eqref{eq:ref5.4} on the \( \psi(\hat{\mathbf{r}},t) \)-function represented by equation \eqref{eq:ref2.1}. We then separate the real and imaginary terms using Eqns. \eqref{eq:refB.6} through \eqref{eq:refB.9} to we arrive at:
\begin{align}
\frac{\hat{\mathbf{p}}_R \cdot \hat{\mathbf{p}}_R}{2m} \psi(\hat{\mathbf{r}},t) &= \frac{(\nabla S)^2}{2m} \psi(\hat{\mathbf{r}},t) \label{eq:refB.11}\\
-\frac{\hat{\mathbf{p}}_I \cdot \hat{\mathbf{p}}_I}{2m} \psi(\hat{\mathbf{r}},t) &= -\frac{\hbar^2}{2m} \frac{\nabla^2 R}{R} \psi(\hat{\mathbf{r}},t) \label{eq:refB.12}\\
\frac{1}{2m} (\hat{\mathbf{p}}_R \cdot \hat{\mathbf{p}}_I) \psi(\hat{\mathbf{r}},t)  &= -\frac{\hbar}{2m} \left( \frac{\nabla R \cdot \nabla S}{R} \right) \psi(\hat{\mathbf{r}},t) \label{eq:refB.13}\\
\frac{1}{2m} (\hat{\mathbf{p}}_I \cdot \hat{\mathbf{p}}_R) \psi(\hat{\mathbf{r}},t) &= -\frac{\hbar}{2m} \left( \frac{\nabla R \cdot \nabla S}{R} \right) \psi(\hat{\mathbf{r}},t)  - \frac{\hbar}{2m} \nabla^2 S \psi(\hat{\mathbf{r}},t) \label{eq:refB.14}\\
e_R(\hat{\mathbf{r}},t) \psi(\hat{\mathbf{r}},t) &= -\frac{\partial S}{\partial t} \psi(\hat{\mathbf{r}},t) \label{eq:refB.15}\\
e_I(\hat{\mathbf{r}},t) \psi(\hat{\mathbf{r}},t) &= \frac{\hbar} {R} \frac{\partial R}{\partial t} \psi(\hat{\mathbf{r}},t) \label{eq:refB.16}
\end{align}
 For Eq. \eqref{eq:ref5.5}, we compute the expectation values after multiplying both sides of Eq. \eqref{eq:ref5.3} by \( 2mR^2 \) and use the partial terms \eqref{eq:refB.13}, \eqref{eq:refB.14}, and \eqref{eq:refB.16}. We invoke the rules to take the divergence of a product of a scalar and a vector. 
 \begin{align} \label{eq:refB.17}
&\int_{-\infty}^{\infty} d\hat{\mathbf{r}} \, |\psi(\hat{\mathbf{r}},t)|^2 \left[ \frac{\partial P}{\partial t} + \nabla \cdot \left(P \frac{\nabla S}{m}\right)\right] =  0 
\end{align}
We follow the following steps for Eq. \eqref{eq:ref5.6}. Eqns. \eqref{eq:refA.3} and \eqref{eq:refB.15} to estimate the ensemble average of total particle energy. 
\begin{align} \label{eq:refB.18}
\langle E \rangle_{\text{ensemble average}} = \langle e_R \rangle & = \int_{-\infty}^{\infty} d\hat{\mathbf{r}} \, |\psi(\hat{\mathbf{r}},t)|^2  \left[-\frac{\partial S}{\partial t} \right].
\end{align}
Thus, the term on the left-hand side of the equation. \eqref{eq:ref2.4}, \( (-\frac{\partial S}{\partial t}) \), is identified with the total energy of the particle. Likewise, the expectation value of \( \frac{p_R \cdot p_R}{2m} \) is given as

\begin{align} \label{eq:refB.19}
\langle \frac{\hat{\mathbf{p}}_R \cdot \hat{\mathbf{p}}_R}{2m} \rangle = \int dr \, \psi^*(\hat{\mathbf{r}},t) \frac{\hat{\mathbf{p}}_R \cdot \hat{\mathbf{p}}_R}{2m} \psi(\hat{\mathbf{r}},t)
= \int d\hat{\mathbf{r}} \, |\psi(\hat{\mathbf{r}},t)|^2 \left[\frac{(\nabla S)^2}{2m}\right]
\end{align}
As shown in Eq. \eqref{eq:refB.4}, the local momentum of a particle is \(  p_R(\hat{\mathbf{r}},t) = \nabla S(\hat{\mathbf{r}},t) \); therefore, its local kinetic energy is \( \frac{(\nabla S)^2}{2m} \). Thus in Bohm's theory, the kinetic energy operator of a particle is \( \frac{\hat{\mathbf{p}}_R \cdot \hat{\mathbf{p}}_R}{2m} \), not the conventional kinetic energy operator given by \( \frac{\hat{\mathbf{p}}_Q \cdot \hat{\mathbf{p}}_Q}{2m} \). To complete the picture, we derive the remaining equations:
\begin{align} 
\langle V(\hat{\mathbf{r}}) \rangle &= \int dr \, \psi^*(\hat{\mathbf{r}},t) V(\hat{\mathbf{r}})\psi(\hat{\mathbf{r}},t)
= \int d\hat{\mathbf{r}} \, |\psi(\hat{\mathbf{r}},t)|^2 \left [V(\hat{\mathbf{r}}) \right ] \label{eq:refB.20} \\
\langle \frac{i\hat{\mathbf{p}}_I \cdot i\hat{\mathbf{p}}_I}{2m} \rangle &= \int dr \, \psi^*(\hat{\mathbf{r}},t) \left [ \frac{i\hat{\mathbf{p}}_I \cdot  i\hat{\mathbf{p}}_I}{2m} \right ] \psi(\hat{\mathbf{r}},t)
= \int d\hat{\mathbf{r}} \, |\psi(\hat{\mathbf{r}},t)|^2 \left [-\frac{\hbar^2}{2m} \frac{\nabla^2 R}{R} \right ] \label{eq:refB.21}
\end{align}
\section*{Acknowledgement}
The author thanks Professors David Crowse and Lawrence Glasser for reviewing the manuscript and providing valuable feedback and encouragement.

\balance
\end{appendices}

\end{document}